\documentclass[%
superscriptaddress,
preprint,
 amsmath,amssymb,
 aps,
prb,
]{revtex4-2}

\usepackage{graphicx}
\usepackage{dcolumn}
\usepackage{bm}
\usepackage[mathlines]{lineno}
\usepackage{mhchem}
\usepackage{xcolor}
\usepackage{ragged2e}
\usepackage{xr}
\begin{document}

\preprint{}

\title{Wrinkle Mediated Phase Transitions in \ce{In2Se3}}

\author{Joseph L. Spellberg}
\thanks{These two authors contributed equally.}
\affiliation
{James Franck Institute, University of Chicago, Chicago, IL 60637, USA}
\affiliation
{Department of Chemistry, University of Chicago, Chicago, IL 60637, USA}

\author{Lina Kodaimati}
\thanks{These two authors contributed equally.}
\affiliation
{Department of Chemistry, University of Chicago, Chicago, IL 60637, USA}

\author{Atreyie Ghosh}
\affiliation
{James Franck Institute, University of Chicago, Chicago, IL 60637, USA}

\author{Prakriti P. Joshi}
\affiliation
{Department of Physical Chemistry, Fritz-Haber-Institut der Max-Planck-Gesellschaft, Berlin 14195, Germany}

\author{Sarah B. King}
\email{sbking@uchicago.edu}
\homepage{http://kinglab.uchicago.edu}
\affiliation
{James Franck Institute, University of Chicago, Chicago, IL 60637, USA}
\affiliation
{Department of Chemistry, University of Chicago, Chicago, IL 60637, USA}

\date{\today}

\begin{abstract}
Crystalline phase transitions in two-dimensional materials enable precise control over electronic and ferroic properties, making them attractive materials for memory and energy storage applications. \ce{In2Se3} is particularly promising because its $\alpha$ and $\beta'$ phases are both stable at room temperature but exhibit distinct ferroic behaviors. However, achieving reliable reversible switching between these states remains challenging. Here, we show that controlled $\beta'\rightarrow\alpha$ phase transitions in 2D \ce{In2Se3} become accessible through laser-induced wrinkling, establishing a room-temperature approach for manipulating ferroic states in \ce{In2Se3} thin films. Combined with thermal annealing for phase recovery, this approach eliminates cryogenic steps and mechanical perturbation while harnessing accumulated internal strain to generate multiphase heterostructures and direct domain reorganization. This pathway for phase transitions in \ce{In2Se3} opens the door for further development in ferroic device architectures and phase-change memory technologies.
\justifying
\end{abstract}

\maketitle

\section{Introduction}

Two-dimensional van der Waals (vdW) materials offer opportunities for next generation devices due to their durability \cite{casillas2015elasticity-681}, tunable electronic properties \cite{dutta2024electronic-bc2}, and ability to be easily integrated into complex frameworks such as sensors, transistors, and photonic circuits when paired with CMOS devices \cite{neumaier2019integrating-112, lemme20222d-f73, romagnoli2018graphene-based-816}. Among the suite of 2D materials, ferroic 2D materials have applications in wide-ranging contexts such as digital memory \cite{lu2024two-dimensional-2b7, xue2022two-dimensional-088}, energy storage \cite{li20232d-693}, and neuromorphic computing \cite{zhai2023reconfigurable-c09,kim20242d-b1d}. Indium(III) selenide (\ce{In2Se3}) is a layered semiconducting 2D material with a number of distinct crystalline phases. At least three phases have reported ferroic properties \cite{manolikas1988new-80e,zhou2017out-of-plane-822, han2024recent-49b,xu2020two-dimensional-5f6,zhang2022atomic-ae3,zhang2024atomic-scale-b78}. The two room temperature stable phases are $\alpha$-\ce{In2Se3}, which exhibits intercorrelated out-of-plane (OOP) and in-plane (IP) distortions \cite{zhou2017out-of-plane-822,cui2018intercorrelated-21f, xue2018roomtemperature-ee3}, and the $\beta'$ phase, containing counterbalancing IP antiparallel distortions \cite{xu2020two-dimensional-5f6,wang2024in-plane-5a7,guo2025femtosecond-4d4}. Both phases can be exfoliated down to the 2D limit and incorporated in layered device architectures \cite{han2024recent-49b}. Furthermore, it is possible to synthesize single flakes and films that contain both phases to develop multiphase heterostructures with regions of distinct ferroic properties \cite{wan2022nonvolatile-ca1, han2023phase-controllable-aea} that can be used for phase change memory applications \cite{wu2023reversible-472}. However, a significant obstacle in realizing such devices is that the $\alpha\rightarrow\beta'$ transition is largely irreversible in thin crystals on substrates \cite{guo2025femtosecond-4d4}. 

The challenge arises from the role of substrate interactions in the stabilization energy of each phase \cite{han2023phase-controllable-aea}. When a thin film or flake of $\alpha$-\ce{In2Se3} is annealed above 200 $^\circ$C it converts to the paraelectric $\beta$ phase. However, upon cooling back to room temperature it does not return to the $\alpha$ phase. Instead the $\beta'$ phase is produced due to strain imparted by the substrate \cite{tao2013crystallinecrystalline-91e,han2023phase-controllable-aea}. One method to recover the $\alpha$ phase is to physically delaminate the film from the substrate, which has been accomplished with mechanical perturbation such as an atomic force microscope (AFM) probe \cite{zheng2022phase-6a7} or in a bending strain cell \cite{han2023phase-controllable-aea}. In these studies, the $\beta'\rightarrow\alpha$ phase transition coincided with the formation of wrinkles in the film, which are known to impact phase transition pathways in exfoliated 2D materials \cite{wu2023reversible-472}. The substrate interaction is critical for stabilizing the $\beta'$ phase and delamination is an effective tool for achieving this phase change. However, these approaches may be difficult to implement \textit{in operando} as they require specific substrates and interfacing with an external instrument to supply the strain.

In this work, we present an approach for the achieving the delamination mediated $\beta'\rightarrow\alpha$ transition via a non-contact and substrate agnostic method, namely wrinkle induction with focused laser illumination. We investigate the efficacy of this approach for cycling between phases, showing that it can be used to repeatedly convert between the two room temperature stable phases.  Furthermore, we examine the role that accumulated strain plays in modifying phase interfaces and domain arrangements. These findings position laser induced wrinkling of \ce{In2Se3} as a new method in the design of 2D material systems for phase change and ferroics applications.

\begin{figure*}
    \centering
    \includegraphics[width=\textwidth]{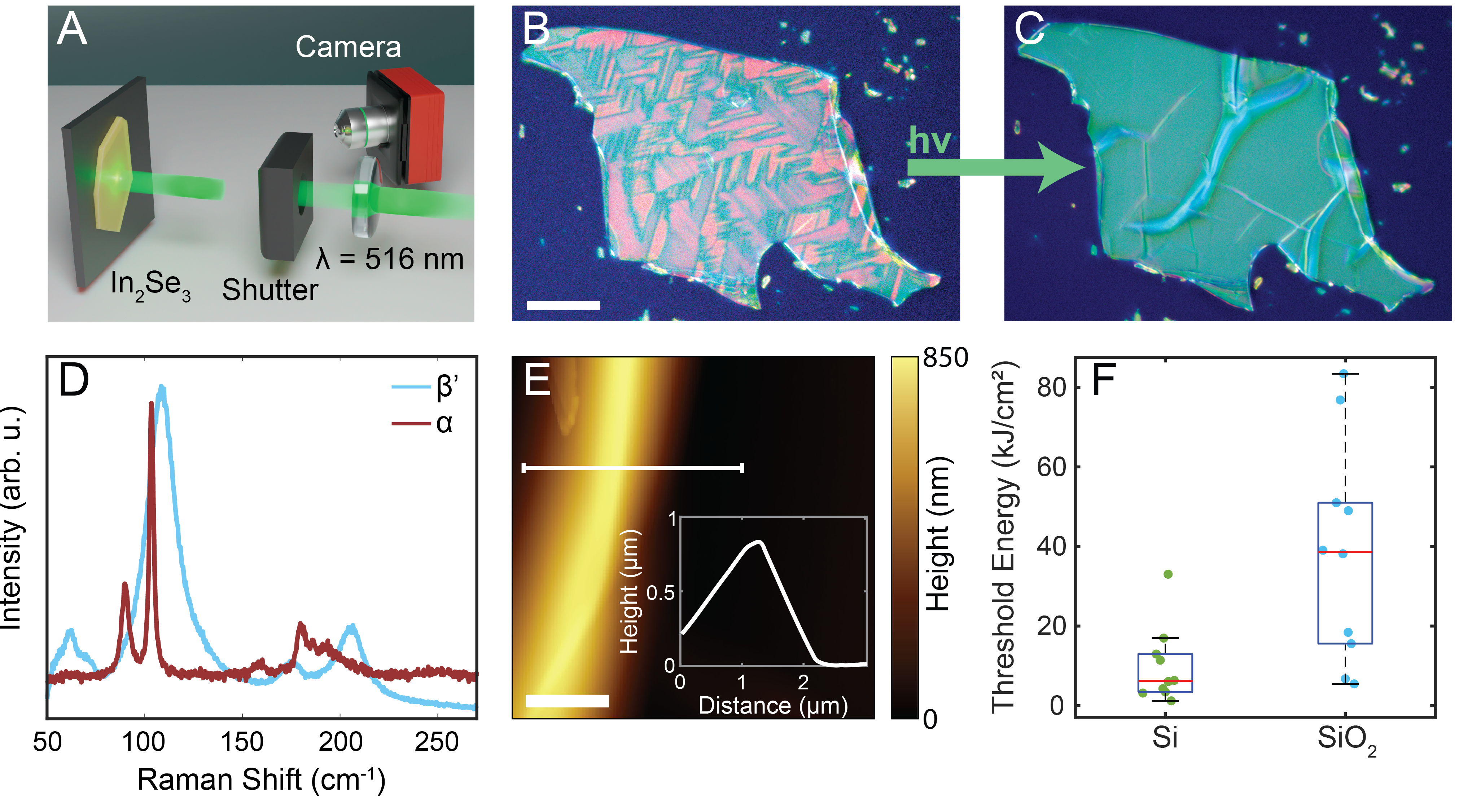}
    \caption{(A) Schematic of the laser induced wrinkle phase transition experiment. The laser is focused onto the sample while being monitored by a camera affixed with a microscope objective. (B,C) Cross-polarized optical images of $\beta'$- and $\alpha$-\ce{In2Se3} before and after laser exposure. Scalebar is $20\ \mu \text{m}$. (D) Raman spectra before and after wrinkling phase transition. Blue line is the pristine flake before wrinkling threshold experiment and red is after wrinkling. (E) AFM image of a wrinkled flake surface. Inset shows height profile along the line cut. Scalebar is $2\ \mu \text{m}$. (F) Box plots of input energy required to induce wrinkles in flakes on Si and \ce{SiO2} substrates. Red lines are medians; boxes and whiskers indicate interquartile ranges. Points are individual measurements.}
    \label{fig:overview}
\end{figure*} 

\section{Results and Discussion}
\subsection*{Phase Transition Pathway}
Figure \ref{fig:overview}(A) shows a schematic representation of our experimental setup for inducing the $\beta'\rightarrow\alpha$ phase transition. A pulsed 516 nm laser focuses on a flake of $\beta'$-\ce{In2Se3} exfoliated on \ce{Si} with either native oxide or thick thermal oxide. A camera mounted with an objective monitors the flake throughout its laser exposure and records the flake's condition in real time. During laser illumination, light scattering prevents us from imaging the flake, so a shutter regularly blocks the beam, enabling imaging. After extended laser exposure, a threshold energy (defined as the total integrated energy from the laser) is reached and the flake rapidly converts to the $\alpha$ phase corresponding to the sudden emergence of wrinkles. The observed wrinkling threshold ranged from about 1-85 kJ/cm$^2$ of total laser energy exposure, depending on the sample. Movie S1 shows a recording of a laser wrinkling experiment at the moment of flake wrinkling when the flake is converted to the $\alpha$ phase. The flake is left under laser exposure for 10s of minutes, but the wrinkling and phase transition occurs rapidly over a few seconds.

Figures \ref{fig:overview}(B,C) show cross-polarized optical microscope images of a flake on \ce{SiO2} substrate before and after laser exposure. A large wrinkle is apparent in the center of the flake after laser exposure that was not present initially. The birefringent contrast in (B) is due to the IP domains in $\beta'$-\ce{In2Se3} which are are not present in (C) $\alpha$-\ce{In2Se3}  \cite{spellberg2024electronic-a46}. Raman spectroscopy confirms these phase assignments (Figure \ref{fig:overview}(D)). Figure \ref{fig:overview}(E) shows an AFM image of the wrinkled flake where the wrinkle is about 800 nm tall and has a bend angle of $44^\circ$, above the critical angle for bend-induced OOP ferroelectric domains \cite{han2023bend-induced-815}.

Figure \ref{fig:overview}(E) shows an overview of the total laser energy required for all wrinkling threshold measurements. More energy is required to induce the phase change on substrates with a 100 nm thick thermal oxide surface than those native oxide which is typically $\sim2$ nm thick \cite{bohling2016self-limitation-681}. Thickness affects the properties of the oxide layer, such as the adhesion strength of adsorbates and hydrophilicity, due to the different structure and composition on the surface as well as surface roughness \cite{williams1974wetting-51b}. This suggests that flakes are held more strongly to the \ce{SiO2} substrates, and more input energy is required for wrinkling and delamination \cite{megra2021enhancement-bb0}. We attribute the wrinkling process in general to strain induced by the mismatch between the thermal expansion coefficients of the flake and substrate as well as nonuniform temperature gradients arising from a beam spot on the same order as the flake size \cite{plechinger2015control-5b5}. Eventually, this results in decoupling at the interface causing delamination and wrinkling. We did not observe a clear trend in wrinkling threshold with respect to total incident laser power, but we did observe that wrinkling would not occur if the beam spot was large relative to the flake size, further suggesting the importance of thermal gradients in the mechanism. Excitation energy dependent Raman spectra do not show distinct electron-phonon coupling in this wavelength range (Figure S1), so we do not attribute the wrinkling to a particular phonon mode of $\ce{In2Se3}$. 

\begin{figure*}
    \centering
    \includegraphics[width=\textwidth]{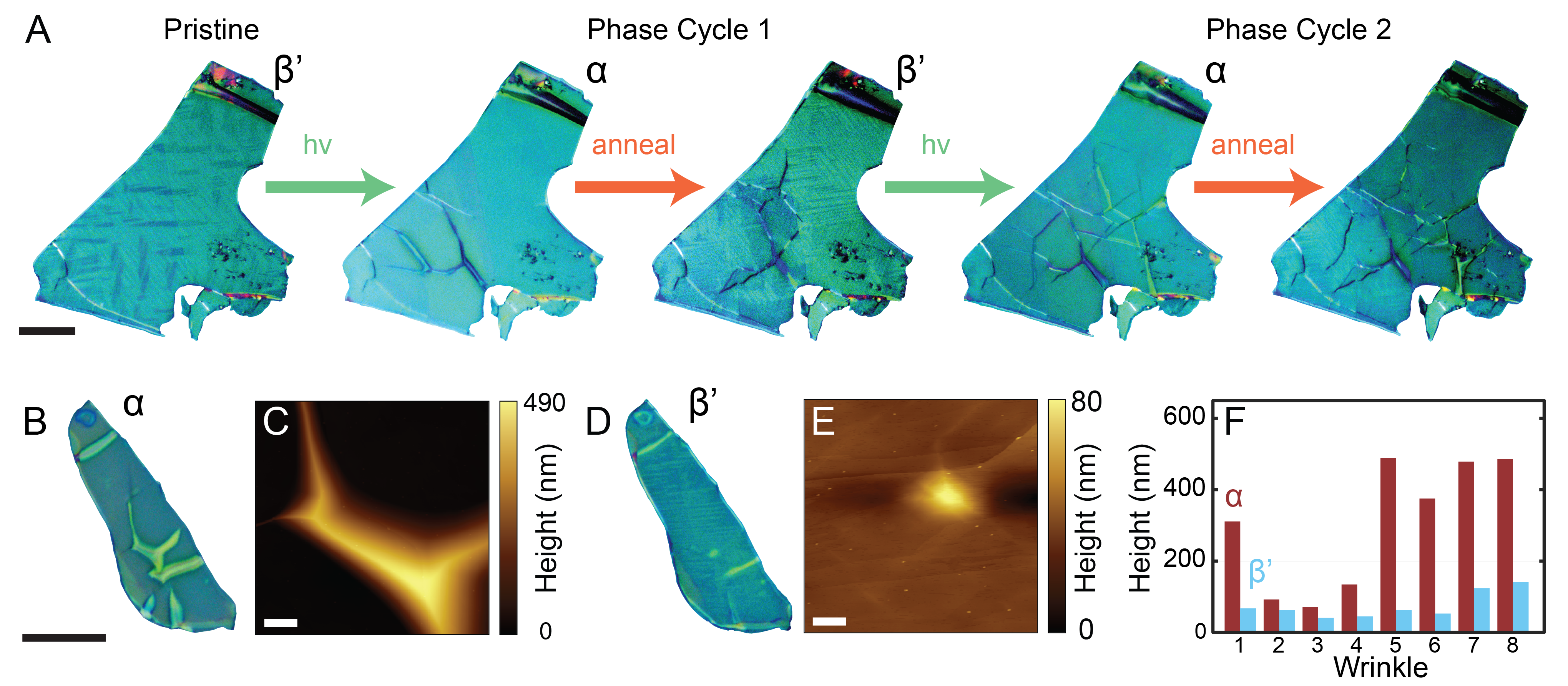}
    \caption{(A) Cross-polarized optical images of an \ce{In2Se3} flake on Si through two phase cycles consisting of laser exposure and annealing. (B,C) Optical and AFM images of a wrinkled $\alpha$ phase flake after laser exposure and (D,E) the same flake after converting back to $\beta'$ via annealing. (F) Heights of various wrinkles before and after annealing measured with AFM . Scalebars are $20\mu$m for optical images and $1\mu$m for AFM images.}
    \label{fig:phase_cycle}
\end{figure*} 

We observed significant variation in the threshold energy between flakes on the same substrate that did not correlate with flake thickness or surface area (Figure S2). This contrasts with expectations for a phase transition mechanism based on simple mechanical sheet bending and delamination \cite{reissner1947bending-0ed,yu2003delamination-019} and the thickness dependence observed in mechanical phase transitions with previous studies \cite{tao2013crystallinecrystalline-91e,zheng2022phase-6a7}. The complex threshold behavior in our experiments arises from the correlated influence of multiple factors in the laser-induced transition, including flake dimensions, substrate contact quality, and local strain fields, all of which can impact the wrinkling dynamics. Importantly, our experiments establish an energy benchmark for this phase transition pathway and highlight the need for future studies using samples with controlled dimensions, such as CVD-grown films \cite{yang2023controllable-354}, to systematically disentangle the parameters governing the wrinkling threshold. Optical images and Raman spectra for all wrinkling experiments in Figure \ref{fig:overview}(F) are provided in Figures S3-S6, with detailed experimental parameters described in the methods section.

Laser-induced wrinkling not only provides a viable pathway for the $\beta'\rightarrow\alpha$ transition but also enables repeatable cycling between phases. Figure \ref{fig:phase_cycle}(A) shows a series of cross-polarized optical images following the same flake through multiple phase changes (associated Raman spectra at each step are shown in Figure S7). The flake is initially in the $\beta'$ phase with a clear domain structure and a relatively flat surface topography. After laser exposure, wrinkles are induced and the domain structures vanish as the flake is converted to $\alpha$-\ce{In2Se3}. When the sample is placed on a hot plate and annealed to $350\ ^\circ$C and then cooled, the domains return but with a new arrangement as the $\beta'$ phase is restored. The wrinkles partially heal, some disappearing entirely, while others are reduced but leave scars on the flake. After a second round of laser exposure, the $\alpha$ phase is once again produced. Some of the previously healed wrinkles return in addition to new wrinkles forming in new locations. Finally, after a second annealing step, the $\beta'$ phase is once again recovered, as evidenced by the reemergence of the birefringent domains (the lack of domains on the right side of the flake will be discussed in the next section). This confirms that our laser-induced wrinkle-mediated strategy is a viable approach for cycling back and forth between these phases repeatedly.  

Figures \ref{fig:phase_cycle}(B,C) show cross-polarized optical and AFM images of a wrinkled $\alpha$ phase flake after laser exposure. There are wrinkles at multiple locations on the flake; the largest is at the center with a height over 400 nm. Figures \ref{fig:phase_cycle}(D,E) show this same flake after annealing and conversion back to the $\beta'$ phase. The optical image shows the IP domains and partial healing of the wrinkles. The wrinkle that was in the center of the flake has almost completely vanished, leaving a remnant scar that is only about 45 nm tall. Examining a collection of several wrinkles across multiple flakes, as shown in Figure \ref{fig:phase_cycle}(F), we find that this behavior is robust; wrinkles always decrease in height after annealing, regardless of their initial height. The wrinkles often decrease in lateral dimension and bend angle as well (Figure S8). This suggests that in addition to inducing the $\alpha\rightarrow\beta'$ phase transition, the thermal energy provided by the annealing step enables the release of bending strain allowing the flake to relax back to a flatter topography. However, we typically do not observe complete wrinkle flattening as there is often a residual structure. The wrinkle annealing process, therefore, permanently alters the strain profile across the flake. 

\subsection*{Multiphase Heterostructures}
\begin{figure}
    \centering
    \includegraphics[width=3.5in]{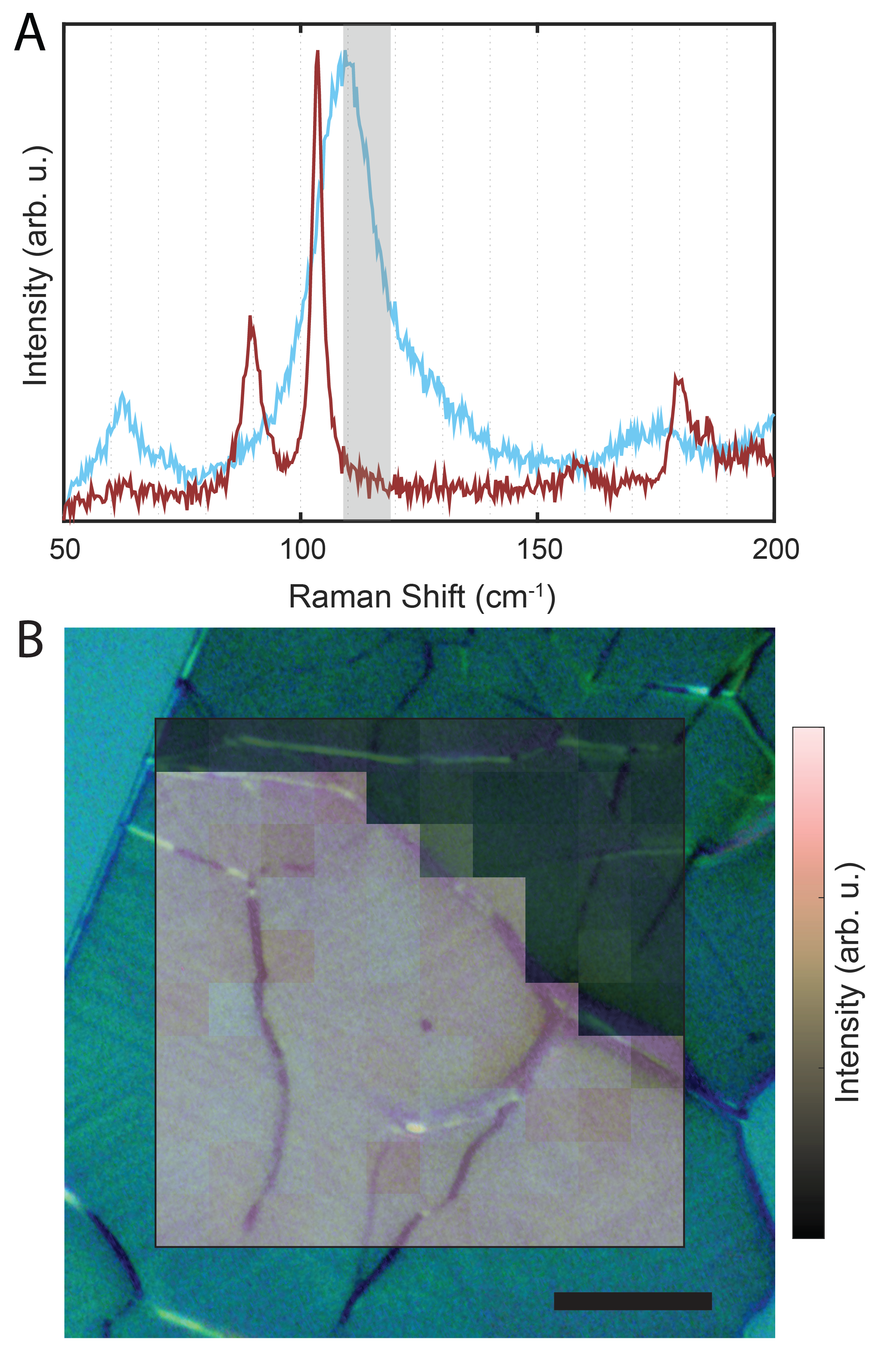}
    \caption{(A) Raman spectra from the right (red trace) and left (blue trace) sides of the flake in Figure \ref{fig:phase_cycle}(A) after phase cycling. (B) Map of integrated Raman intensity for the range (109-119 cm$^{-1}$) highlighted in gray in (A) overlayed over a cross-polarized optical image of the flake. Pink corresponds to greater intensity of the $\beta'$-\ce{In2Se3} $A_1$ mode. Scalebar is $10\ \mu$m.}
    \label{fig:Raman_map}
\end{figure}  
After two rounds of phase cycling, the flake investigated in Figure \ref{fig:phase_cycle}(A) only exhibited domains on the left side of the flake while the right side was much darker in the cross-polarized optical image. Raman spectra taken on these two regions, Figure \ref{fig:Raman_map}(B), show that the left side of the flake converted to $\beta'$ after the final annealing step while the right side remained in the $\alpha$ phase. Figure \ref{fig:Raman_map}(B) shows a map of Raman intensity between 109-119 cm$^{-1}$ overlaid on a cross-polarized optical image of the flake. This spectral region between 109-119 cm$^{-1}$ has high intensity from the $\beta'$-\ce{In2Se3} $A_1$ mode but is spectrally resolved from the modes in the $\alpha$ phase. Therefore, the Raman map provides a local measure of crystalline phase \cite{balakrishnan2018epitaxial-eda}. The left section of the flake, where domains are observed, is $\beta'$-\ce{In2Se3} while the right side is entirely $\alpha$ phase. The boundary between these regions follows closely along a wrinkle that formed during the laser exposure.

Over the course of repeated wrinkling and healing, the structural distortion in that location became pronounced, such that the phase change could not propagate past it in the final annealing step, leading to the formation of a lateral multiphase heterostructure. Such heterostructures can also be produced when the laser is focused smaller than the entire flake and contained within a small region bounded by a topographic feature such as a step edge (Figure S9). There, the wrinkling is confined to the step-edge bounded region and the rest of the flake remains in the $\beta'$ after laser exposure. We also observe that prolonged high power laser exposure can generate a film of $\gamma$-\ce{In2Se3} on top of the flake while the bulk of the flake is preserved \cite{marsillac1996experimental-92a}. To our knowledge, this is the first report of these three phases of \ce{In2Se3} being present simultaneously in the same flake.

The boundaries between the $\alpha$ and $\beta'$ regions form sharp lateral junctions. Devices incorporating such junctions would benefit from the intrinsic ferroelectric polarization in \ce{In2Se3} as a means for using external biasing to modulate switching behavior \cite{han2023phase-controllable-aea,xiong2023out-of-plane-3d3}. The approach illustrated here, utilizing intraflake strain to generate phase boundaries, is a compelling avenue for the continued research in heterostructure development.

\subsection{Domain Arrangements}
\begin{figure*}
    \centering
    \includegraphics[width=\textwidth]{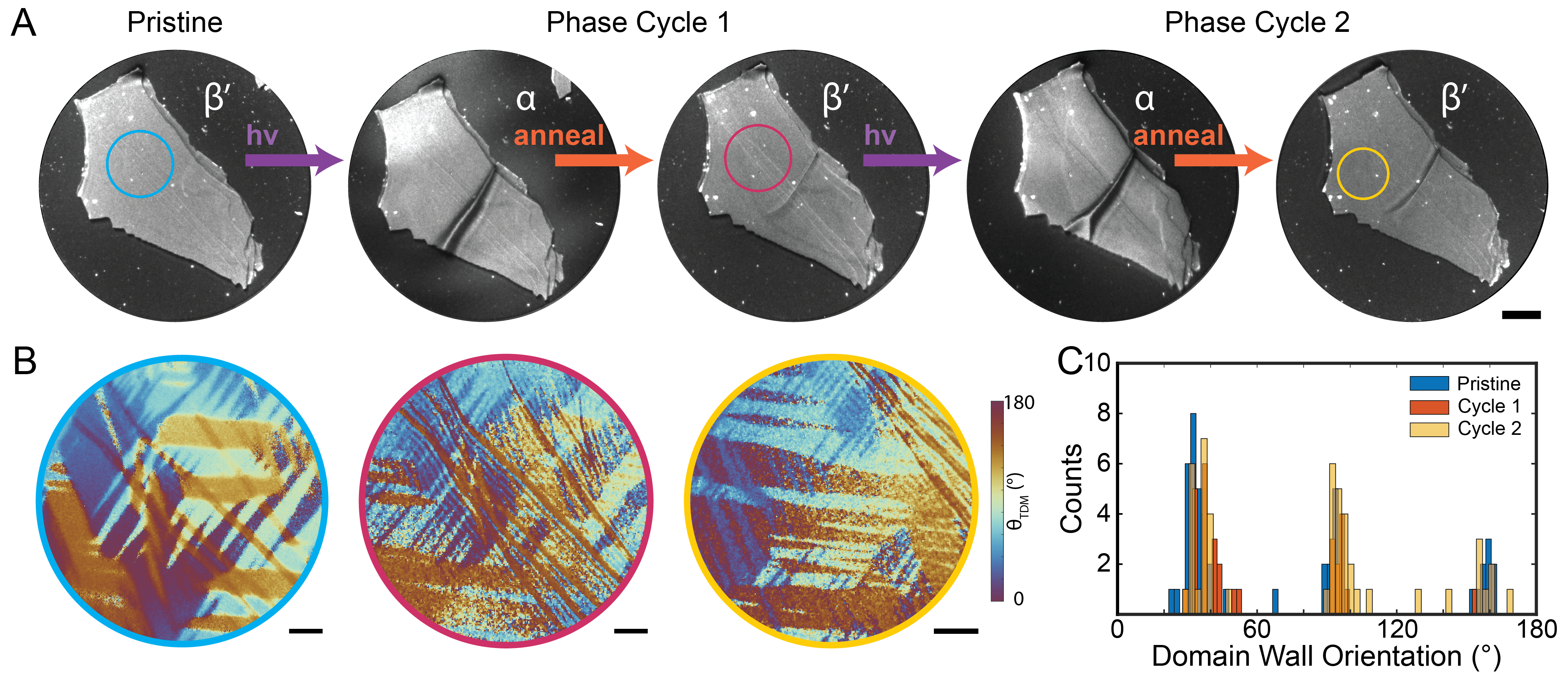}
    \caption{(a) Static PEEM images of \ce{In2Se3} through two phase cycles consisting of laser ($h\nu=3.1$ eV) exposure and annealing. Scalebar is $10\mu$m. (B) Maps of $\theta_\text{TDM}$ at from the marked regions in (A) for each step in the phase cycling process. Scalebar is $2\mu$m. (C) Histogram of domain boundary orientations throughout the phase cycling process.}
    \label{fig:PD_PEEM}
\end{figure*} 

Finally, we investigate the impact of the wrinkle annealing procedure on the arrangement of IP domains in $\beta'$-\ce{In2Se3}. We have previously shown that these domains can be imaged with polarization-dependent photoemission electron microscopy (PD-PEEM) which achieves greater spatial resolution and orientation fidelity than is possible with cross-polarized optical microscopy \cite{spellberg2024electronic-a46, ghosh2024polarization-dependent-27a}. We therefore employ this technique to observe the domain arrangement of a flake at each step in the phase cycling process. Figure \ref{fig:PD_PEEM}(A) shows static PEEM images of a flake  through two wrinkle-anneal cycles. After laser exposure, a single wrinkle forms across the middle of the flake which is then mostly healed upon annealing. After a second laser exposure, the wrinkle returns in the same location and is slightly more prominent. At each step in the process when the flake was in the $\beta'$ phase a PD-PEEM experiment was conducted to image the domain arrangement. Details of this method have been discussed previously \cite{neff2017imaging-719,spellberg2024electronic-a46,joshi2022nanoimaging-30e,ghosh2024polarization-dependent-27a}. In brief, linearly polarized laser excitation impinges on the sample at near-normal incidence. When the electric field of the light is in parallel to the local transition dipole moment of the material, photoemission is maximized. In a material with IP domains such as $\beta'$-\ce{In2Se3}, the transition dipole moment orientation, $\theta_\text{TDM}$, differs between domains due to the rotated optical axes. This causes a spatially dependent phase shift in the $\theta_\text{TDM}$ photoemission response, allowing different domains to be resolved. The color scale \cite{crameri2020misuse-075,noauthornoyearcrameri-0dc} in Figure \ref{fig:PD_PEEM}(B) maps $\theta_\text{TDM}$ to reveal the arrangement of the domain structure in the regions indicated in Figure \ref{fig:PD_PEEM} (A). The maps are generated by plotting the phase shift of the photoemission response at each location when fit to $I(\theta_E)=A\cos^2(\theta_E-\theta_\text{TDM})+C$, where $\theta_E$ is the orientation of the laser polarization. Coefficient of determination, $R^2$, for these fits are shown in Figure S10. 

As the flake is cycled between the phases, the domains return in different locations and arrangements. Previous work showed that when heating to the high temperature $\beta$ phase and cooling back to $\beta'$, domains do not change shape and position \cite{zheng2018room-25a}. Here, however, we observe a change in the domain arrangement due to modification of the flake's internal strain environment because of the wrinkle mediated phase transition. This change may stem from the substrate-flake interaction where the flake has a changed contact area with the substrate as well as intraflake effects at, and around, the wrinkle scars. Domains in $\beta'$-\ce{In2Se3} are known to be sensitive to applied stress \cite{xu2021two-dimensional-9b9} and local strain gradients have been shown to modify domain arrangements and create new domain boundaries in $\alpha$-\ce{In2Se3} \cite{han2023bend-induced-815,he2024ferroelectric-4e1}. Therefore, we attribute the modified domain arrangement to the new strain environment present in the flake after phase cycling. Domain and domain boundary arrangements play an important role in the observable physical properties of a ferroic material and can dictate how applications may be realized \cite{fousek2001domain-de6, yang2010above-bandgap-8ab,guzelturk2023subnanosecond-82f}. As such, domain reorganization due to wrinkling strain in $\beta'$-\ce{In2Se3} could be advantageous for a variety of applications ranging from digital memory, to photonics and energy storage \cite{li2008controlled-444,seidel2009conduction-749,ma2019atomically-779,xu2025antiferroelectric-898}.
 
We note that the region on which we imaged the domain structure also contained topographical step edges, which act as a coupling structure from which fringes can propagate. These may be diffraction effects \cite{chelaru2006fringe-394} or possibly waveguide modes due to the high positive value of the real part of the $\beta'$-\ce{In2Se3} dielectric function at visible wavelengths \cite{rieger2025imaging-3c4,chen2024characteristic-6cf}. They are unlikely to be plasmonic in nature due to the dielectric function remaining positive across this entire range, however, additional work is needed to fully understand the fringing behavior. The fringes partially obscure the domain structure that we intend to observe \cite{chen2024characteristic-6cf}. As such, the final $\theta_\text{TDM}$ is taken at a slightly different location. This location still overlaps spatially with the originally mapped location and we can verify that the domains are distinct from the structure in both of the first two images.

The orientations of domain boundaries in $\beta'$-\ce{In2Se3} are controlled by the crystalline axes of the parent $\beta$ phase structure. Specifically, they can only form along the [11$\bar{2}$0] or [1$\bar{1}$00] directions. This results in six possible orientations that a domain wall can take which are conserved across the 10s of $\mu m$ spatial scale of individual flakes \cite{xu2021two-dimensional-9b9,spellberg2024electronic-a46}. Measuring the domain wall orientations in the domain maps throughout the phase cycling process we observe that they form along the same directions with respect to real space at each step, Figures \ref{fig:PD_PEEM}(C) and S11. Interestingly, in this flake we observe three predominant domain boundary orientations spaced by $60^\circ$, indicating that on this particular flake the domain boundaries formed along primarily one set of lattice vectors. While domain shapes and arrangement are changed, the parent lattice orientation is conserved through the phase cycling process. This suggests that during laser induced wrinkling there is no intermediate amorphous phase and that the process is largely a displacive phase transition \cite{modi2024electrically-4de,salje1991crystallography-18a}.

\section{Conclusion}
In summary, we present laser induced wrinkling as a new pathway for converting $\beta'\rightarrow\alpha$-\ce{In2Se3} in a non-contact, substrate agnostic manner. This phase transition is reversible upon annealing allowing it to be repeated multiple times, thereby completing the phase transition cycle between the two phases without the need for a cryogenic step or other apparatus \cite{wu2023reversible-472}. The repeated wrinkling and healing modifies the topography and strain profile across the flake which can be used to create multiphase heterostructures within a single flake and modify the spontaneous arrangement of domains in the $\beta'$ phase. This finding positions wrinkle induced strain as a powerful tool in the development of applications ranging from heterostructure junctions to domain patterning for information storage. The robust nature of the parent crystalline axes of the material is also encouraging for the use of this process in phase change memory applications \cite{yoo2023review-f21}. In order to meet the ever increasing demands on our energy and data storage infrastructures, innovative and novel device design paradigms are needed and 2D materials such as \ce{In2Se3} are well suited to this goal. The phase transition pathways and ability to modify mesoscale structure shown here offer possible opportunities for addressing these challenges.

\section{Methods}
\subsection{Sample Preparation}
Bulk $\alpha$-\ce{In2Se3} crystals grown by chemical vapor transport were purchased from 2D Semiconductors and mechanically exfoliated in \ce{N2} atmosphere. A polydimethylsiloxane stamp was used to transfer flakes onto n-doped Si with native oxide or 100 nm thermal oxide. Samples were annealed to $350^\circ$C for 30 min in \ce{N2} atmosphere to produce $\beta'$-\ce{In2Se3}. The same annealing procedure was used for phase-cycled samples. The timescale for oxidation of \ce{In2Se3} flakes in air is on the order of days to weeks, much longer than the amount of cumulative air exposure during the experiments in this study \cite{yan2023phase-621}. However, we observe that flakes tend to be more sensitive to degradation during exfoliation and annealing so these steps where performed in \ce{N2} atmosphere out of caution.

\subsection{Wrinkling Threshold}
The $\beta'\rightarrow\alpha$-\ce{In2Se3} phase transition was achieved by using laser illumination in ambient conditions. The laser ($\lambda=516\pm2\space\text{nm}$, repetition rate = 4 MHz) set-up is pictured in Figure 1(A). A recording was collected for the duration of the experiment to determine the time at which wrinkling occurs in a given flake. A shutter was in place to allow for the viewing of the flake throughout the duration of the experiment; all experiments were conducted with the shutter open for 1.5 s for every 1 s closed. Observed wrinkling times ranged from around 1 minute to up to 20 minutes when accounting for the time spent with the shutter closed. The power of the laser and the beam spot size were used to calculate the wrinkling threshold energy for the flake which is defined as the time integrated total energy per area supplied to the flake. The beam spot diameters were typically around 200 $\mu$m. For the phase-cycling experiments, the laser-wrinkling maintains the same set-up. Cross-polarized optical images and Raman microscopy provided phase confirmation before and after wrinkling. Contrast is boosted in cross-polarized optical images to more clearly visualize the domains.

\subsection{Atomic Force Microscopy}
Atomic force microscopy (Bruker MultiMode8, tapping mode and ScanAsyst) was used to determine flake thickness and topography. AFM images were corrected in Gwyddion where the minimum point was set to zero. Other methods of correction included row alignment and data leveling.
\subsection{Raman Spectroscopy}
Raman microscopy (HORIBA LabRAM HR Evolution confocal Raman microscope) was used to determine sample phase. All measurements were taken with a 633 nm laser unless noted otherwise. 
\subsection{Photoemission Electron Microscopy}
PEEM experiments were conducted in ultrahigh vacuum conditions (base pressure $1\times10^{-10}$ mbar) in a photoelectron microscope from Focus GmbH. Static PEEM images were collected via single photon photoemission with a broadband Hg arc lamp as the illumination source. Polarization dependent PEEM measurments were conducted with the second harmonic of a homebuilt OPA with photon energy $h\nu=3.1$eV ($\lambda = 401 \pm5$ nm, repetition rate = 4MHz). Fluence was about 100 $\mu J/\text{cm}^2$, the low power was used to avoid wrinkling the flakes during the course of the PEEM experiment. Flakes were wrinkled \textit{in situ} with the same excitation source. For the first wrinkling step the fluence was about 6000 $\mu J/\text{cm}^2$ and the wrinkling phase change occurred after a few hours of exposure. Due to the lower wrinkling threshold in the latter phase cycle, the second wrinkling occurred after a few hours under about 100 $\mu J/\text{cm}^2$. Between phase cycle steps the sample was transferred back into the glovebox for annealing.

In generating the $\theta_\text{TDM}$ maps, to improve signal quality difference images and spatial binning were used in accordance with previously reported methods \cite{spellberg2024electronic-a46}. In the second and third measurement the wrinkling threshold is lower due to the remnant scar so a lower fluence was used which resulted in a lower signal to noise ratio. Therefore, $4\times4$ binning was also used for the latter two measurements.

\section{Data and Code Availability}

\begin{acknowledgments}
This work was funded by the Office of Basic Energy Sciences, U.S. Department of Energy (Grant No. DE-SC0021950). We acknowledge the use of UChicago Materials Research Science and Engineering Center shared facilities funded by the NSF under award number DMR-2011854. We acknowledge Ryo Mizuta Graphics for providing the Blender assets used in some of the figures in this publication.
\end{acknowledgments}

\section{References}
\bibliography{references3}

\begin{figure*}[h]
    \centering
    \includegraphics[]{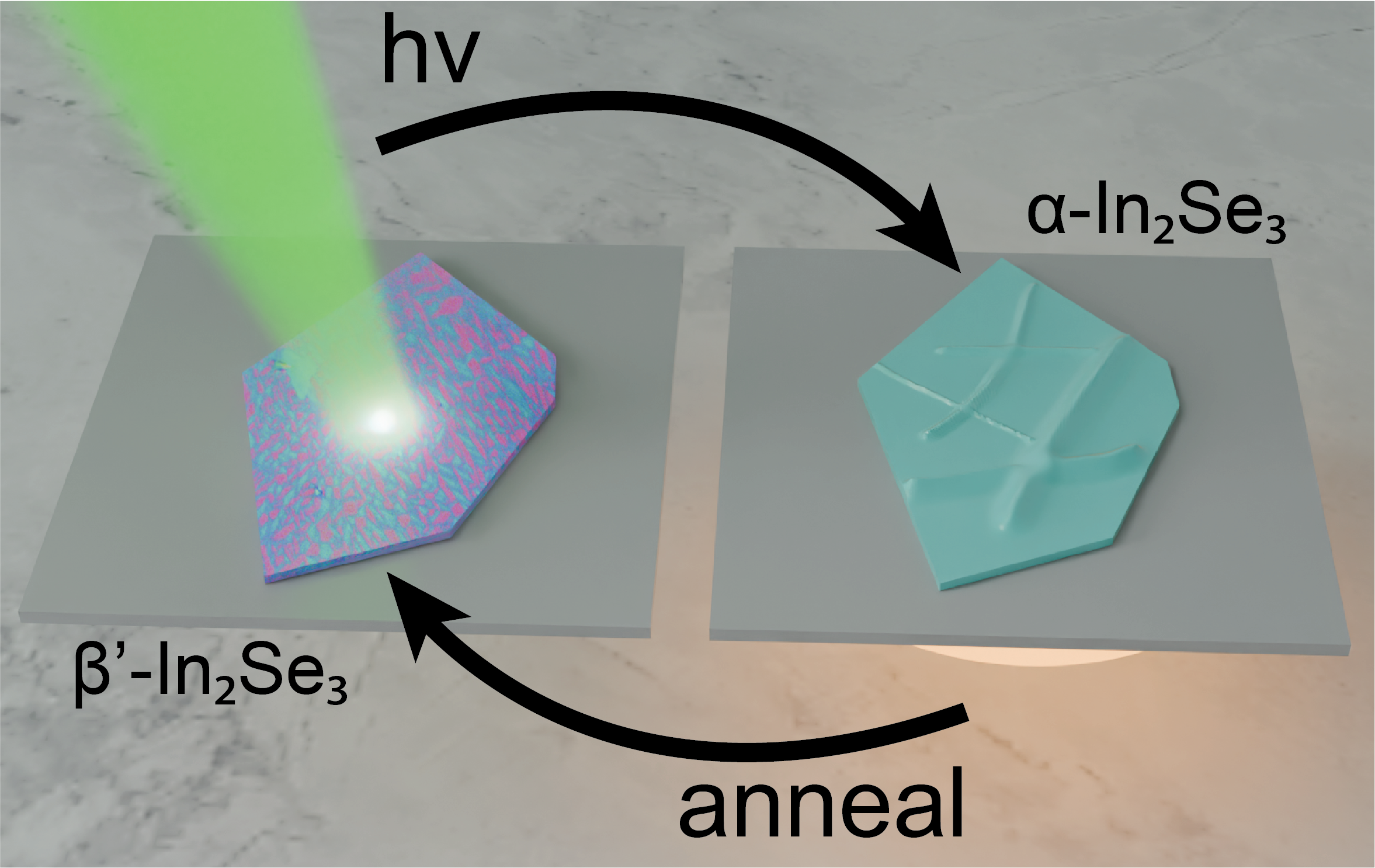}
    \caption{Table of Contents Figure}
\end{figure*}

\end{document}